\documentclass[prl,reprint,superscriptaddress]{revtex4-2}
\usepackage{amsmath,moreverb,graphics,graphicx,color,amsfonts,amssymb,bbold}
\usepackage{hyperref}
\usepackage[shortlabels]{enumitem}
\usepackage{multirow}
\usepackage[normalem]{ulem}

\newcommand{\beq}{\begin{equation}}
\newcommand{\eeq}{\end{equation}}

\newcommand{\be}{\mathcal{E}}
\newcommand{\gmn}{g^{\mu\nu}}
\newcommand{\avg}[1]{\overline{#1}}
\newcommand{\fluc}[1]{\widetilde{#1}}
\newcommand{\au}{\avg{u}}
\newcommand{\ab}{\avg{b}}
\newcommand{\fu}{\fluc{u}}
\newcommand{\fb}{\fluc{b}}
\newcommand{\azp}{\avg{z}_+}
\newcommand{\azm}{\avg{z}_-}
\newcommand{\fzp}{\fluc{z}_+}
\newcommand{\fzm}{\fluc{z}_-}

\begin{document}
\title{Exact nonlinear solutions for three-dimensional Alfv\'en-wave packets in relativistic magnetohydrodynamics}
\author{Alfred Mallet}
\affiliation{Space Sciences Laboratory, University of California, Berkeley CA 94720, USA}
\author{Benjamin D. G. Chandran}
\affiliation{Space Science Center, University of New Hampshire, Durham, NH 03824, USA}
\date{\today}
\begin{abstract}
We show that large-amplitude, non-planar, Alfv\'en wave (AW) packets are exact nonlinear solutions of the relativistic MHD equations when the total magnetic-field strength in the local fluid rest frame ($b$) is a constant. We
derive analytic expressions relating the components of the fluctuating velocity and magnetic field. We also show that these constant-$b$ AWs propagate without distortion at the relativistic Alfv\'en velocity and never steepen into shocks. These findings and the observed abundance of large-amplitude, constant-$b$ AWs in the solar wind suggest that such waves may be present in relativistic outflows around compact astrophysical objects.
\end{abstract}
\maketitle
\paragraph{Introduction.---}
Black holes and neutron stars are among the most remarkable objects in the universe. In addition to warping spacetime, they generate powerful plasma outflows, which, in the case of supermassive black holes, can manifest as radio sources that extend up to a million light years through intergalactic space. Because
the plasma in these outflows is often relativistic, it is
of interest to study how the basic building blocks of plasma physics, for example the plasma waves, behave in a relativistic system. 

One of the most important waves in non-relativistic plasma physics is the \uline{Alfv\'en Wave} (AW) \citep{alfven1942,barnes1971,goldstein1974,barnes1974}. This wave has prompted a great deal of study,
in part because of its ubiquitous presence in spacecraft observations of the solar wind
\citep{belcher1971}. The prevalence of this wave in the solar wind 
may be due to the fact that long-wavelength propagating fluctuations that are not AWs
quickly dissipate, either through steepening into shocks, turbulent mixing, or damping due to wave-particle interactions
\citep{barnes1974,cohen1974,vasquez1996a,schekochihin2016}. In contrast, AWs are undamped in a collisionless plasma in the long-wavelength limit \citep{barnes1971}, and they undergo only weak turbulent mixing when most of the AWs propagate in a single direction along the background magnetic field lines  (as happens in the solar wind, in which most of the AWs propagate away from the Sun in the plasma rest frame). AWs in non-relativistic plasmas also
possess a polarization state in which the waves do not steepen into shocks, irrespective of their amplitude. This is the `spherical polarization state,' in which the total magnetic-field strength is a constant. Indeed, in a homogeneous, non-relativistic plasma, a nonlinear, three-dimensional AW packet in which the total magnetic-field strength, density, and pressure are constant is an exact solution to the compressible magnetohydrodynamic (MHD) equations \cite{goldstein1974}. In the solar wind, the observed AWs are often nearly perfectly spherically polarized.

Alfv\'en waves play an important role in space and astrophysical plasmas. For example, they contribute substantially to the heating of the solar corona and the energization of the solar wind.  Convective  motions at the solar photosphere shake the magnetic field lines that connect the solar surface to the distant interplanetary medium, thereby launching AWs that transport energy outward from the solar surface. In many models, the dissipation of this AW energy flux is the dominant heating mechanism in the solar corona and solar wind \citep{cranmer2007,verdini2009,vanderholst2014,chandran2021}. A similar energization mechanism could arise in relativistic astrophysical plasmas, in which a dense central object (e.g., a black-hole accretion disk, or the surface of a proto-neutron star) has a turbulent surface and is threaded by a magnetic field (e.g., \citet{metzger2007}).

AWs also play a crucial role in the transport and confinement of cosmic rays. When the average cosmic-ray drift velocity through a plasma exceeds the the Alfv\'en speed, the AW becomes unstable and grows, leading to wave pitch-angle scattering of the cosmic rays \citep{lerche1966,wentzel1968,kulsrud1969}. This same process plays a critical role in diffusive shock acceleration. The streaming of cosmic rays away from a shock in the upstream direction amplifies AWs, which scatter the cosmic rays, causing them to return to the shock, thereby enabling the repeated shock crossings required to accelerate particles to high energies \citep{bell1978}.

The tendency for AWs to develop spherical polarization  in non-relativistic plasmas has important implications for the way that AWs affect the transport of energetic particles and energize 
plasma outflows. For example, in contrast to large-amplitude, linearly polarized AWs, 
large-amplitude spherically polarized AWs do not cause magnetic mirroring of cosmic rays.
In addition, numerical simulations suggest that 
when the amplitudes of the fluctuating and background magnetic fields are comparable, 
spherically polarized AWs necessarily develop discontinuous magnetic-field rotations \cite{valentini2019, squire2020,shoda2021}. Copious abrupt magnetic-field rotations are indeed observed in the solar wind close to the Sun \cite{kasper2019,bale2019,horbury2020}, but fewer are observed farther away, implying that these discontinuities erode over time, possibly via plasma instabilities \cite{tenerani2020}. The development and decay of these discontinuities provide a dissipation channel for AWs that can alter the rate at which wave energy is thermalized and, in principle, the way that the dissipated wave energy is apportioned among different particle species (c.f. \citep{howes2010,kawazura2020}). If a relativistic analog of the spherically polarized state exists for the relativistic AW,
this could have important implications for energetic particle propagation and turbulent heating in relativistic plasmas. This possibility is the focus of this Letter.

Previous work on large-amplitude relativistic AWs has been limited to the so-called `simple wave', in which the magnetic-field strength in the local fluid frame is a constant and the fluctuations depend only on a single scalar variable $\phi(x^\mu)$ \citep{greco1972,barnes1971,anile1989}. It was shown that the simple AW propagates without steepening.
This was apparently rediscovered by \citet{heyvaerts2012}, who also showed that the simple AWs are necessarily planar (1+1 dimensional).

In this Letter, we extend this work to more general fluctuations, without assuming plane polarization. We show that any fluctuations in the magnetic-field four-vector and velocity four-vector that are proportional to each other in the same way as linear AWs are exact nonlinear solutions to the relativistic MHD equations in flat spacetime, provided that the 
mass density, internal energy,  pressure, and background magnetic field are constants. In these solutions, the magnetic-field strength in the local fluid rest frame is a constant.
The resulting wave packets propagate through the plasma at
the relativistic Alfv\'en velocity without steepening into shocks. 

\paragraph{Elsasser formulation of GRMHD.---}
The equations of GRMHD \citep{anile1989,gammie2003} describe the motion of a perfectly conducting fluid under the influence of the electromagnetic fields and gravity, and may be derived assuming that the electric field vanishes in the local fluid rest frame. These equations are, first, the conservation of mass
\beq
\nabla_\nu (\rho u^\nu) = 0,\label{eq:consmass}
\eeq
the stress-energy equation 
\beq
\nabla_\nu T^{\mu\nu} = 0,\label{eq:consse}
\eeq
and the induction equation
\beq
\nabla_\nu \left( b^\mu u^\nu - b^\nu u^\mu \right)=0.\label{eq:ind}
\eeq
In these equations, $\nabla_\nu$ denotes the covariant derivative, $\rho$ is the mass density, $u^\mu$ is the fluid 4-velocity, the GRMHD stress-energy tensor is
\beq
T^{\mu\nu} = \be u^\mu u^\nu  - b^\mu b^\nu + \left( p+\frac{b^2}{2}\right)\gmn,\label{eq:se}
\eeq
where $\gmn$ is the metric tensor, the magnetic-field 4-vector is
\beq
b^\mu = \frac{1}{2} \epsilon^{\mu\nu\kappa\lambda}u_\nu F_{\lambda\kappa},\label{eq:4b}
\eeq
with $b^2 = b^\mu b_\mu>0$, $F_{\lambda\kappa}$ the Faraday tensor divided by $\sqrt{4\pi}$, $\epsilon^{\mu\nu\kappa\lambda}$ the Levi-Civita tensor, and 
\beq
\be = \rho + U + p + b^2,
\eeq
where $U$ is the internal energy and $p$ is the thermal pressure. We use units such that the speed of light $c=1$. We use the notation
\beq
A^2 = A^\mu A_\mu 
\eeq
to denote the magnitude squared of any 4-vector $A^\mu$; for spacelike 4-vectors, we also write $A=\sqrt{A^2}$.

The 4-velocity satisfies
\beq
u^2 = -1,\label{eq:uconstraint}
\eeq
and Equation (\ref{eq:4b}) implies that
\beq
u_\mu b^\mu = 0.\label{eq:ubconstraint}
\eeq
\citet{chandran2018} noticed that, just as in non-relativistic MHD, (\ref{eq:consse}) and (\ref{eq:ind}) may be cast in a useful pseudo-symmetric Elsasser \citep{elsasser1950} form by multiplying (\ref{eq:ind}) by $
\pm\be^{1/2}$, adding to (\ref{eq:consse}), and dividing the two resulting equations by $
\be$. This results in
\beq
\nabla_\nu \left( z_\pm^\mu z_\mp^\nu + \Pi \gmn\right) + \left( \frac{3}{4}z_\pm^\mu z_\mp^\nu + \frac{1}{4} z_\mp^\mu z_\pm^\nu + \Pi \gmn\right) \frac{\partial_\nu \be}{\be},\label{eq:els}
\eeq
where
\beq
z_\pm^\mu = u^\mu \mp  \frac{b^\mu}{\be^{1/2}}, \quad \Pi = \frac{2p + b^2}{2\be},\label{eq:elsdef}
\eeq
and $\partial_\nu$ refers to differentiation with respect to the coordinate $\nu$. Eqs.~(\ref{eq:els}), along with Eq.~(\ref{eq:consmass}) and an equation of state, comprise the Elsasser formulation of GRMHD.

In the following, we restrict ourselves to the case of special relativity, for which the Minkowski metric may be written in Cartesian coordinates $g^{\mu\nu}=\mathrm{diag}(-1,1,1,1)$, and the covariant derivative reduces to the simpler 4-gradient operator $\nabla_\nu = \partial_\nu$. 

\paragraph{Fluctuations on a uniform background.---}
We take each quantity to be the sum of a background value plus a fluctuation:
\[
\rho = \bar{\rho} + \tilde{\rho} \qquad 
p = \bar{p} + \tilde{p} \qquad
U = \bar{U} + \tilde{U}
\]
\beq
u^\mu = \bar{u}^\mu + \fu^\mu \qquad b^\mu = \bar{b}^\mu + \fb^\mu \qquad
z_\pm^\mu = \bar{z}_\pm^\mu + \tilde{z}_\pm^\mu
\eeq
We take the background quantities to be uniform in space and time,
\beq
\{ \bar{\rho},\, \bar{\be}, \,\bar{\Pi},\, \bar{u}^\mu, \,\bar{b}^\mu \} \,= \;{\rm constants},\label{eq:bg}
\eeq
and consider fluctuations satisfying 
\beq
\tilde{\rho} = 
\tilde{\be} = 
\tilde{\Pi} = 0, \qquad
\fu^\mu = - \frac{\fb^\mu}{{\cal E}^{1/2}}.
\label{eq:fluctuations}
\eeq
The final equation of (\ref{eq:fluctuations}) implies that
$\fzm^\mu = 0$. We assume that the fluctuations are localised in spacetime around a sequence of events $X^\mu$; at another sequence of events ${X'}^\mu$, $b^\nu(X'^\mu) \to \ab^\mu$ and $u^\mu(X'^\mu) \to\au^\mu$ as $|(X-X')^2|\to\infty$.

It follows from (\ref{eq:uconstraint}) that
\beq
u^\mu u_\mu = \au^2 + \fu^2 + 2 \au^\mu \fu_\mu = -1.\label{eq:totu}
\eeq
The spacetime localization of the fluctuations combined with the constancy of~$\bar{u}^\mu$ implies that
$\au^2=-1$, and hence
\beq
\fu^2 = - 2 \au^\mu \fu_\mu.\label{eq:flucconstraintau}
\eeq
Equation (\ref{eq:ubconstraint}) further restricts the solution by requiring that
\beq
\frac{1}{\be^{1/2}}u^\mu b_\mu = \frac{1}{\be^{1/2}}\au^\mu \ab_\mu - \fu^2 - \azp^\mu\fu_\mu = 0.
\label{eq:ubzero}
\eeq 
The localization of the fluctuations and the constancy of $\bar{u}^\mu$ and $\bar{b}^\mu$ imply that $\au^\mu\ab_\mu =0$, so (\ref{eq:ubzero}) becomes
\beq
\fu^2 = -\azp^\mu\fu_\mu.\label{eq:flucconstraintazp}
\eeq
Subtracting (\ref{eq:flucconstraintau}) from twice (\ref{eq:flucconstraintazp}), we find that
\beq
\fu^2 = \frac{2}{\be^{1/2}}\ab^\mu \fu_\mu.\label{eq:flucconstraintab}
\eeq


Finally, we calculate the scalar $b^2$. This is
\beq
b^2=b^\mu b_\mu = \ab^2 + \be\fu^2 - 2\be^{1/2}\ab^\mu \fu_\mu = \ab^2,
\eeq
a constant, where we have used (\ref{eq:flucconstraintab}) in the last equality. Thus, the wave packet has constant 4-magnetic-field magnitude, analogous to the constant-$B^2$ constraint for a large-amplitude AW in non-relativistic MHD \citep{goldstein1974,barnes1974}. 

At each point in space-time, we may boost into an accelerating frame moving with the instantaneous local fluid velocity $u^\mu$, the \uline{local fluid rest frame} (LFRF). In this frame, $b^t=0$ and, therefore, the magnetic-field three-vector has magnitude-squared $B^2=b^2$, which is a constant and therefore the same at each point. $B^2$ is not spatially constant in any fixed inertial frame.

Eq.~(\ref{eq:consmass}) with $\rho$ constant and flat space-time gives
\beq
\partial_\nu \fu^\nu = \partial_\nu \fzp^\nu =0,\label{eq:divfree}
\eeq
and the $+$ Elsasser equation (\ref{eq:els}) then gives
\beq
\azm^\nu \partial_\nu \fzp^\mu = 0, \label{eq:prop}
\eeq
with the $-$ Elsasser equation vanishing by virtue of (\ref{eq:divfree}) and (\ref{eq:fluctuations}). Eq.~(\ref{eq:prop}) is a \uline{linear} wave equation for the evolution of $\fzp^\mu$; thus, a three-dimensional Alfv\'enic wavepacket of (apparently; see Eq.~\ref{eq:limit}) arbitrary amplitude and arbitrary shape propagates without distortion on a homogeneous background.
\paragraph{Components in the background rest frame.---}
We define a \uline{background rest frame} (BRF) \footnote{In \citet{chandran2018}, this was called the \uline{average fluid rest frame} (AFRF).} in which the homogeneous background (\ref{eq:bg}) is at rest. In this frame $\au^\mu = (1,0,0,0)$, and $\ab^\mu = (0, 0,0,b)$, where we have chosen to align the $z$ direction with the background magnetic field. Since $u^\mu$ is a future-directed four-velocity, it is straightforward to show, working in the BRF and using (\ref{eq:totu}), that $\fu^2 \geq 0$, a relation that holds in all frames since $\fu^2$ is a scalar. (The equality $\fu^2 = 0$ is obtained only when $\fu^\mu = 0$.)
Calculating (\ref{eq:flucconstraintau}) and (\ref{eq:flucconstraintab}) in the BRF, the $t$ and $z$ components of the fluctuation are given by
\beq
\fu^t = \frac{1}{2}\fu^2, \quad 
\fu^z = - \frac{\be^{1/2}}{2 b} \fu^2\quad \text{(BRF)}.\label{eq:futz}
\eeq
and the magnitude of the remaining (perpendicular) fluctuation components $\fu_\perp=\sqrt{(\fu^x)^2 + (\fu^y)^2}$ is
\begin{align}
\fu_\perp = \sqrt{\fu^2 + (\fu^t)^2 - (\fu^z)^2}
=\fu\sqrt{1-\frac{\fu^2}{4\sigma^2}}\quad \text{(BRF)},\label{eq:fuperp}
\end{align}
where $\sigma^2 = b^2/(\rho+U+p)$. Providing $\fu^2$ thus gives us nearly all the information in the fluctuation four-vector, apart from the direction in the $y$-$z$ plane that the perpendicular fluctuation points. To determine this, we must use (\ref{eq:divfree}) and a particular functional form for $\fu^2$.

Evaluating the fluid 3-velocity $v^i=u^i/u^t$ and magnetic-field 3-vector $B^i=b^i u^t - b^t u^i$, it is clear that in the BRF, $B^2$ is not constant, the 3-vector magnetic and velocity fluctuations are not related to each other by a constant of proportionality, and do not in general even point in the same direction.

\paragraph{Maximum amplitude.---}
Eq.~(\ref{eq:fuperp}) implies an upper limit on the magnitude of the fluctuations, 
\beq
\fu \leq \fu_{\rm max} = \frac{2 b}{\sqrt{\rho+U + p}}.\label{eq:limit}
\eeq
In the non-relativistic case, this has been recently noticed in solar wind AWs by \citep{matteini2018}. The observed magnitude of the fluid 3-velocity in the BRF is
\beq
v=\frac{\sqrt{u^iu_i}}{u^t} = 
\frac{\fu\sqrt{1+\fu^2/4}}{1+\fu^2/2} \quad\text{(BRF)},
\eeq
an increasing function of $\fu$; $v<1$ and $v\to 1$ as $\fu\to\infty$.

\paragraph{Alfv\'en velocity and wave frame.---}
The propagation of the wave is controlled by the constant timelike 4-vector $\azm^\mu$, with 
\beq
\azm^2 = \frac{b^2}{\be} - 1 <0. 
\eeq
(In the limit $b^2\to\infty$ while keeping $\rho$, $U$, and $p$ constant, $\azm^2 \to 0$.) An observer moving with 4-velocity $\azm^\mu / \sqrt{-\azm^2}$ sees a time-independent structure; such an observer is in the \uline{wave frame} (WF). The WF 3-velocity relative to another frame is
$v_w^i = \azm^i/\azm^t$, 
and specifically, relative to the BRF is
\beq
v_w^i = v_A^i = \frac{b}{\be^{1/2}} (0,0,1) \quad\text{(BRF)}.\label{eq:va}
\eeq
Thus, the 3D relativistic AW propagates along the background field lines at this relativistic Alfv\'en velocity, just like the planar AW \citep{barnes1971,heyvaerts2012}. Note that this does not depend on $\fu^\mu$, so the wave does not steepen into a shock. As $b^2 \to \infty$, $v_A^2 \to 1$, while if all of $b^2/\rho,U/\rho,p/\rho\ll 1$, we recover the usual non-relativistic Alfv\'en velocity.

\paragraph{Structure in the wave frame.---}
In the WF, the spatial components of $z_-^\mu = \azm^\mu$ are zero and so \beq
u^i = - b^i/\be^{1/2}\quad \text{(WF)}. \label{eq:wfprop}
\eeq
Using (\ref{eq:ubconstraint}), we may obtain
\beq
b^t = -\be^{1/2} \gamma v^2 \quad \text{(WF)}.\label{eq:btwf}
\eeq
where $v^2$ is the square of the three-velocity $v^i=u^i/\gamma$ and $\gamma=u^t$. Then, we may calculate the magnetic field 3-vector,
\beq
B^i = b^i u^t - b^t u^i = -\be^{1/2} v^i \quad \text{(WF)},\label{eq:Biwf}
\eeq
so in the WF the 3-velocity is parallel and proportional to the magnetic field 3-vector, just like in the non-relativistic case. This also implies $\partial_i v^i=0$ in the WF. We may also calculate 
\beq
b^2 = b^i b_i - (b^t)^2 = \be v^2=B^2 \quad\text{(WF)},\label{eq:b2wf}
\eeq
and so, since $b^2$ is constant, in the WF $v^2$ and $B^2$ are both constant, just like in the non-relativistic case \citep{matteini2015}.

Let us consider the components of the stress-energy tensor (\ref{eq:se}) in the WF. First, using (\ref{eq:btwf}) and (\ref{eq:b2wf}),
\begin{align}
T^{tt} = \rho+U+\frac{3b^2}{2}\quad\text{(WF)},
\end{align}
a space-time constant for our solution. Applying (\ref{eq:btwf}) again,
\beq
T^{ti}=T^{it}= \be v^i\quad\text{(WF)},
\eeq
which from (\ref{eq:Biwf}) has no spatial divergence, maintaining the constancy of $T^{tt}$ in the equation $\partial_\nu T^{t\nu} = 0$. Finally,
\beq
T^{ij} = \left( p+ \frac{b^2}{2}\right)\delta^{ij}\quad\text{(WF)},
\eeq
a space-time constant, thus enforcing the constancy in time of $T^{it}$ in the equations $\partial_\nu T^{i\nu} = 0$. The cancellation of the first two terms in the space-space components of (\ref{eq:se}) in the WF generalizes the result for the nonrelativistic AW that the centrifugal force exactly balances the tension force in the magnetic field, keeping the fluid flowing exactly along the field lines in the WF.

\paragraph{Discussion.---}
Our analysis has shown that some of the unique properties of the AW survive, even with relativistic fluctuation velocities, and arbitrarily strong magnetic field strength. Specifically, just as in the non-relativistic case \citep{goldstein1974}, a three-dimensional Alfv\'enic structure propagates in time without steepening into a shock, no matter its fluctuation amplitude; equivalently, the propagation velocity in the rest frame of the background is always the relativistic Alfv\'en velocity (\ref{eq:va}), which is independent of the fluctuation amplitude. Also analogous to the non-relativistic case, the magnitude $b^2$ of the magnetic field 4-vector $b^\mu$ is a space-time constant. This implies correlations between different components of the fluctuation to enforce this constraint. Unlike in the non-relativistic case, in a general inertial frame the magnetic-field 3-vector does not have constant magnitude; however, in the wave frame moving at the Alfv\'en velocity, both the velocity and magnetic field 3-vectors have constant magnitude, as in the non-relativistic case. Also in the wave frame, the plasma 3-velocity is parallel and proportional to the magnetic-field 3-vector.

In what situations might one see large-amplitude relativistic AWs? In a statistically homogeneous medium, one might expect equal fluxes of $\fluc{z}_\pm^\mu$ AWs, a nonlinear, turbulent situation. However, if the waves are excited by some particular event or set of events, they will mainly travel away from that event. We might postulate (inspired by non-relativistic plasma physics) that sufficiently far from the source, the other, non-Alfv\'enic modes largely dissipate, and then we are left with just the AWs. This situation would be relevant, for example, in outflows around a compact object like a black hole \citep{chandran2018}. One caveat is that in this case the background is likely to be highly inhomogeneous, and these inhomogeneities will reflect the waves and thus drive turbulence. However, as previously mentioned, in the non-relativistic case it can be shown that even including this turbulence \citep{cranmer2005,verdini2007,ballegooijen2016,ballegooijen2017,perezchandran2013,chandran2019}, in fact the normalized amplitude $\fluc{B}/\avg{B}$ of the primary outward-travelling AWs tends to grow with distance from the central object (in the solar wind case, the Sun \citep{parker1965,hollweg1974}), with the other, reflected components remaining relatively small: i.e., the fluctuations are approximately large-amplitude AWs. This is thought to be a possible origin for the ``switchbacks"; abrupt magnetic-field reversals recently observed by NASA's Parker Solar Probe in the near-Sun solar wind \citep{squire2020}. If this result carries over to the relativistic case (as it appears to, cf. \citet{chandran2018}), large-amplitude AWs of the type discussed in this Letter may also exist around astrophysical compact objects, and may therefore need to be taken into account when studying the heating and observed dynamics of the surrounding magnetized plasma \citep{akiyama2021,akiyama2021b}.
\acknowledgements
AM was supported by NASA grant 80NSSC21K0462 and NASA contract NNN06AA01C. BDGC was supported in part by NASA grants NNX17AI18G and 80NSSC19K0829.
\bibliography{mainbib2}

\begin{thebibliography}{45}%
\makeatletter
\providecommand \@ifxundefined [1]{%
 \@ifx{#1\undefined}
}%
\providecommand \@ifnum [1]{%
 \ifnum #1\expandafter \@firstoftwo
 \else \expandafter \@secondoftwo
 \fi
}%
\providecommand \@ifx [1]{%
 \ifx #1\expandafter \@firstoftwo
 \else \expandafter \@secondoftwo
 \fi
}%
\providecommand \natexlab [1]{#1}%
\providecommand \enquote  [1]{``#1''}%
\providecommand \bibnamefont  [1]{#1}%
\providecommand \bibfnamefont [1]{#1}%
\providecommand \citenamefont [1]{#1}%
\providecommand \href@noop [0]{\@secondoftwo}%
\providecommand \href [0]{\begingroup \@sanitize@url \@href}%
\providecommand \@href[1]{\@@startlink{#1}\@@href}%
\providecommand \@@href[1]{\endgroup#1\@@endlink}%
\providecommand \@sanitize@url [0]{\catcode `\\12\catcode `\$12\catcode
  `\&12\catcode `\#12\catcode `\^12\catcode `\_12\catcode `\%12\relax}%
\providecommand \@@startlink[1]{}%
\providecommand \@@endlink[0]{}%
\providecommand \url  [0]{\begingroup\@sanitize@url \@url }%
\providecommand \@url [1]{\endgroup\@href {#1}{\urlprefix }}%
\providecommand \urlprefix  [0]{URL }%
\providecommand \Eprint [0]{\href }%
\providecommand \doibase [0]{https://doi.org/}%
\providecommand \selectlanguage [0]{\@gobble}%
\providecommand \bibinfo  [0]{\@secondoftwo}%
\providecommand \bibfield  [0]{\@secondoftwo}%
\providecommand \translation [1]{[#1]}%
\providecommand \BibitemOpen [0]{}%
\providecommand \bibitemStop [0]{}%
\providecommand \bibitemNoStop [0]{.\EOS\space}%
\providecommand \EOS [0]{\spacefactor3000\relax}%
\providecommand \BibitemShut  [1]{\csname bibitem#1\endcsname}%
\let\auto@bib@innerbib\@empty
\bibitem [{\citenamefont {Alfv{\'e}n}(1942)}]{alfven1942}%
  \BibitemOpen
  \bibfield  {author} {\bibinfo {author} {\bibfnamefont {H.}~\bibnamefont
  {Alfv{\'e}n}},\ }\href@noop {} {\bibfield  {journal} {\bibinfo  {journal}
  {Nature}\ }\textbf {\bibinfo {volume} {150}},\ \bibinfo {pages} {405}
  (\bibinfo {year} {1942})}\BibitemShut {NoStop}%
\bibitem [{\citenamefont {Barnes}\ and\ \citenamefont
  {Suffolk}(1971)}]{barnes1971}%
  \BibitemOpen
  \bibfield  {author} {\bibinfo {author} {\bibfnamefont {A.}~\bibnamefont
  {Barnes}}\ and\ \bibinfo {author} {\bibfnamefont {G.~C.}\ \bibnamefont
  {Suffolk}},\ }\href@noop {} {\bibfield  {journal} {\bibinfo  {journal} {J.
  Plasma Phys.}\ }\textbf {\bibinfo {volume} {5}},\ \bibinfo {pages} {315}
  (\bibinfo {year} {1971})}\BibitemShut {NoStop}%
\bibitem [{\citenamefont {Goldstein}\ \emph {et~al.}(1974)\citenamefont
  {Goldstein}, \citenamefont {Klimas},\ and\ \citenamefont
  {Barish}}]{goldstein1974}%
  \BibitemOpen
  \bibfield  {author} {\bibinfo {author} {\bibfnamefont {M.~L.}\ \bibnamefont
  {Goldstein}}, \bibinfo {author} {\bibfnamefont {A.}~\bibnamefont {Klimas}},\
  and\ \bibinfo {author} {\bibfnamefont {F.}~\bibnamefont {Barish}},\
  }\href@noop {} {\bibfield  {journal} {\bibinfo  {journal} {Solar Wind III}\
  ,\ \bibinfo {pages} {385}} (\bibinfo {year} {1974})}\BibitemShut {NoStop}%
\bibitem [{\citenamefont {Barnes}\ and\ \citenamefont
  {Hollweg}(1974)}]{barnes1974}%
  \BibitemOpen
  \bibfield  {author} {\bibinfo {author} {\bibfnamefont {A.}~\bibnamefont
  {Barnes}}\ and\ \bibinfo {author} {\bibfnamefont {J.~V.}\ \bibnamefont
  {Hollweg}},\ }\href@noop {} {\bibfield  {journal} {\bibinfo  {journal} {J.
  Geophys. Res.}\ }\textbf {\bibinfo {volume} {79}},\ \bibinfo {pages} {2302}
  (\bibinfo {year} {1974})}\BibitemShut {NoStop}%
\bibitem [{\citenamefont {{Belcher}}\ and\ \citenamefont
  {{Davis}}(1971)}]{belcher1971}%
  \BibitemOpen
  \bibfield  {author} {\bibinfo {author} {\bibfnamefont {J.~W.}\ \bibnamefont
  {{Belcher}}}\ and\ \bibinfo {author} {\bibfnamefont {L.}~\bibnamefont
  {{Davis}}, \bibfnamefont {Jr.}},\ }\href
  {https://doi.org/10.1029/JA076i016p03534} {\bibfield  {journal} {\bibinfo
  {journal} {J. Geophys. Res.}\ }\textbf {\bibinfo {volume} {76}},\ \bibinfo
  {pages} {3534} (\bibinfo {year} {1971})}\BibitemShut {NoStop}%
\bibitem [{\citenamefont {Cohen}\ and\ \citenamefont
  {Kulsrud}(1974)}]{cohen1974}%
  \BibitemOpen
  \bibfield  {author} {\bibinfo {author} {\bibfnamefont {R.~H.}\ \bibnamefont
  {Cohen}}\ and\ \bibinfo {author} {\bibfnamefont {R.~M.}\ \bibnamefont
  {Kulsrud}},\ }\href@noop {} {\bibfield  {journal} {\bibinfo  {journal} {Phys.
  Fluids}\ }\textbf {\bibinfo {volume} {17}},\ \bibinfo {pages} {2215}
  (\bibinfo {year} {1974})}\BibitemShut {NoStop}%
\bibitem [{\citenamefont {Vasquez}\ and\ \citenamefont
  {Hollweg}(1996)}]{vasquez1996a}%
  \BibitemOpen
  \bibfield  {author} {\bibinfo {author} {\bibfnamefont {B.~J.}\ \bibnamefont
  {Vasquez}}\ and\ \bibinfo {author} {\bibfnamefont {J.~V.}\ \bibnamefont
  {Hollweg}},\ }\href@noop {} {\bibfield  {journal} {\bibinfo  {journal} {J.
  Geophys. Res. Space Phys.}\ }\textbf {\bibinfo {volume} {101}},\ \bibinfo
  {pages} {13527} (\bibinfo {year} {1996})}\BibitemShut {NoStop}%
\bibitem [{\citenamefont {{Schekochihin}}\ \emph {et~al.}(2016)\citenamefont
  {{Schekochihin}}, \citenamefont {{Parker}}, \citenamefont {{Highcock}},
  \citenamefont {{Dellar}}, \citenamefont {{Dorland}},\ and\ \citenamefont
  {{Hammett}}}]{schekochihin2016}%
  \BibitemOpen
  \bibfield  {author} {\bibinfo {author} {\bibfnamefont {A.~A.}\ \bibnamefont
  {{Schekochihin}}}, \bibinfo {author} {\bibfnamefont {J.~T.}\ \bibnamefont
  {{Parker}}}, \bibinfo {author} {\bibfnamefont {E.~G.}\ \bibnamefont
  {{Highcock}}}, \bibinfo {author} {\bibfnamefont {P.~J.}\ \bibnamefont
  {{Dellar}}}, \bibinfo {author} {\bibfnamefont {W.}~\bibnamefont
  {{Dorland}}},\ and\ \bibinfo {author} {\bibfnamefont {G.~W.}\ \bibnamefont
  {{Hammett}}},\ }\href {https://doi.org/10.1017/S0022377816000374} {\bibfield
  {journal} {\bibinfo  {journal} {J. Plasma Phys.}\ }\textbf {\bibinfo {volume}
  {82}},\ \bibinfo {pages} {905820212} (\bibinfo {year} {2016})}\BibitemShut
  {NoStop}%
\bibitem [{\citenamefont {Cranmer}\ \emph {et~al.}(2007)\citenamefont
  {Cranmer}, \citenamefont {Van~Ballegooijen},\ and\ \citenamefont
  {Edgar}}]{cranmer2007}%
  \BibitemOpen
  \bibfield  {author} {\bibinfo {author} {\bibfnamefont {S.~R.}\ \bibnamefont
  {Cranmer}}, \bibinfo {author} {\bibfnamefont {A.~A.}\ \bibnamefont
  {Van~Ballegooijen}},\ and\ \bibinfo {author} {\bibfnamefont {R.~J.}\
  \bibnamefont {Edgar}},\ }\href@noop {} {\bibfield  {journal} {\bibinfo
  {journal} {Astrophys. J. Suppl. Ser.}\ }\textbf {\bibinfo {volume} {171}},\
  \bibinfo {pages} {520} (\bibinfo {year} {2007})}\BibitemShut {NoStop}%
\bibitem [{\citenamefont {Verdini}\ \emph {et~al.}(2009)\citenamefont
  {Verdini}, \citenamefont {Velli}, \citenamefont {Matthaeus}, \citenamefont
  {Oughton},\ and\ \citenamefont {Dmitruk}}]{verdini2009}%
  \BibitemOpen
  \bibfield  {author} {\bibinfo {author} {\bibfnamefont {A.}~\bibnamefont
  {Verdini}}, \bibinfo {author} {\bibfnamefont {M.}~\bibnamefont {Velli}},
  \bibinfo {author} {\bibfnamefont {W.~H.}\ \bibnamefont {Matthaeus}}, \bibinfo
  {author} {\bibfnamefont {S.}~\bibnamefont {Oughton}},\ and\ \bibinfo {author}
  {\bibfnamefont {P.}~\bibnamefont {Dmitruk}},\ }\href@noop {} {\bibfield
  {journal} {\bibinfo  {journal} {Astrophys. J. Lett.}\ }\textbf {\bibinfo
  {volume} {708}},\ \bibinfo {pages} {L116} (\bibinfo {year}
  {2009})}\BibitemShut {NoStop}%
\bibitem [{\citenamefont {van~der Holst}\ \emph {et~al.}(2014)\citenamefont
  {van~der Holst}, \citenamefont {Sokolov}, \citenamefont {Meng}, \citenamefont
  {Jin}, \citenamefont {Manchester~IV}, \citenamefont {Toth},\ and\
  \citenamefont {Gombosi}}]{vanderholst2014}%
  \BibitemOpen
  \bibfield  {author} {\bibinfo {author} {\bibfnamefont {B.}~\bibnamefont
  {van~der Holst}}, \bibinfo {author} {\bibfnamefont {I.~V.}\ \bibnamefont
  {Sokolov}}, \bibinfo {author} {\bibfnamefont {X.}~\bibnamefont {Meng}},
  \bibinfo {author} {\bibfnamefont {M.}~\bibnamefont {Jin}}, \bibinfo {author}
  {\bibfnamefont {W.~B.}\ \bibnamefont {Manchester~IV}}, \bibinfo {author}
  {\bibfnamefont {G.}~\bibnamefont {Toth}},\ and\ \bibinfo {author}
  {\bibfnamefont {T.~I.}\ \bibnamefont {Gombosi}},\ }\href@noop {} {\bibfield
  {journal} {\bibinfo  {journal} {The Astrophysical Journal}\ }\textbf
  {\bibinfo {volume} {782}},\ \bibinfo {pages} {81} (\bibinfo {year}
  {2014})}\BibitemShut {NoStop}%
\bibitem [{\citenamefont {Chandran}(2021)}]{chandran2021}%
  \BibitemOpen
  \bibfield  {author} {\bibinfo {author} {\bibfnamefont {B.~D.}\ \bibnamefont
  {Chandran}},\ }\href@noop {} {\bibfield  {journal} {\bibinfo  {journal}
  {arXiv preprint arXiv:2101.04156}\ } (\bibinfo {year} {2021})}\BibitemShut
  {NoStop}%
\bibitem [{\citenamefont {Metzger}\ \emph {et~al.}(2007)\citenamefont
  {Metzger}, \citenamefont {Thompson},\ and\ \citenamefont
  {Quataert}}]{metzger2007}%
  \BibitemOpen
  \bibfield  {author} {\bibinfo {author} {\bibfnamefont {B.~D.}\ \bibnamefont
  {Metzger}}, \bibinfo {author} {\bibfnamefont {T.~A.}\ \bibnamefont
  {Thompson}},\ and\ \bibinfo {author} {\bibfnamefont {E.}~\bibnamefont
  {Quataert}},\ }\href@noop {} {\bibfield  {journal} {\bibinfo  {journal}
  {Astrophys. J.}\ }\textbf {\bibinfo {volume} {659}},\ \bibinfo {pages} {561}
  (\bibinfo {year} {2007})}\BibitemShut {NoStop}%
\bibitem [{\citenamefont {Lerche}(1966)}]{lerche1966}%
  \BibitemOpen
  \bibfield  {author} {\bibinfo {author} {\bibfnamefont {I.}~\bibnamefont
  {Lerche}},\ }\href@noop {} {\bibfield  {journal} {\bibinfo  {journal} {Phys.
  Fluids}\ }\textbf {\bibinfo {volume} {9}},\ \bibinfo {pages} {1073} (\bibinfo
  {year} {1966})}\BibitemShut {NoStop}%
\bibitem [{\citenamefont {Wentzel}(1968)}]{wentzel1968}%
  \BibitemOpen
  \bibfield  {author} {\bibinfo {author} {\bibfnamefont {D.~G.}\ \bibnamefont
  {Wentzel}},\ }\href@noop {} {\bibfield  {journal} {\bibinfo  {journal}
  {Astrophys. J.}\ }\textbf {\bibinfo {volume} {152}},\ \bibinfo {pages} {987}
  (\bibinfo {year} {1968})}\BibitemShut {NoStop}%
\bibitem [{\citenamefont {Kulsrud}\ and\ \citenamefont
  {Pearce}(1969)}]{kulsrud1969}%
  \BibitemOpen
  \bibfield  {author} {\bibinfo {author} {\bibfnamefont {R.}~\bibnamefont
  {Kulsrud}}\ and\ \bibinfo {author} {\bibfnamefont {W.~P.}\ \bibnamefont
  {Pearce}},\ }\href@noop {} {\bibfield  {journal} {\bibinfo  {journal}
  {Astrophys. J.}\ }\textbf {\bibinfo {volume} {156}},\ \bibinfo {pages} {445}
  (\bibinfo {year} {1969})}\BibitemShut {NoStop}%
\bibitem [{\citenamefont {Bell}(1978)}]{bell1978}%
  \BibitemOpen
  \bibfield  {author} {\bibinfo {author} {\bibfnamefont {A.}~\bibnamefont
  {Bell}},\ }\href@noop {} {\bibfield  {journal} {\bibinfo  {journal} {MNRAS}\
  }\textbf {\bibinfo {volume} {182}},\ \bibinfo {pages} {443} (\bibinfo {year}
  {1978})}\BibitemShut {NoStop}%
\bibitem [{\citenamefont {Valentini}\ \emph {et~al.}(2019)\citenamefont
  {Valentini}, \citenamefont {Malara}, \citenamefont {Sorriso-Valvo},
  \citenamefont {Bruno},\ and\ \citenamefont {Primavera}}]{valentini2019}%
  \BibitemOpen
  \bibfield  {author} {\bibinfo {author} {\bibfnamefont {F.}~\bibnamefont
  {Valentini}}, \bibinfo {author} {\bibfnamefont {F.}~\bibnamefont {Malara}},
  \bibinfo {author} {\bibfnamefont {L.}~\bibnamefont {Sorriso-Valvo}}, \bibinfo
  {author} {\bibfnamefont {R.}~\bibnamefont {Bruno}},\ and\ \bibinfo {author}
  {\bibfnamefont {L.}~\bibnamefont {Primavera}},\ }\href@noop {} {\bibfield
  {journal} {\bibinfo  {journal} {Astrophys. J. Lett.}\ }\textbf {\bibinfo
  {volume} {881}},\ \bibinfo {pages} {L5} (\bibinfo {year} {2019})}\BibitemShut
  {NoStop}%
\bibitem [{\citenamefont {Squire}\ \emph {et~al.}(2020)\citenamefont {Squire},
  \citenamefont {Chandran},\ and\ \citenamefont {Meyrand}}]{squire2020}%
  \BibitemOpen
  \bibfield  {author} {\bibinfo {author} {\bibfnamefont {J.}~\bibnamefont
  {Squire}}, \bibinfo {author} {\bibfnamefont {B.~D.}\ \bibnamefont
  {Chandran}},\ and\ \bibinfo {author} {\bibfnamefont {R.}~\bibnamefont
  {Meyrand}},\ }\href@noop {} {\bibfield  {journal} {\bibinfo  {journal}
  {Astrophys. J. Lett.}\ }\textbf {\bibinfo {volume} {891}},\ \bibinfo {pages}
  {L2} (\bibinfo {year} {2020})}\BibitemShut {NoStop}%
\bibitem [{\citenamefont {Shoda}\ \emph {et~al.}(2021)\citenamefont {Shoda},
  \citenamefont {Chandran},\ and\ \citenamefont {Cranmer}}]{shoda2021}%
  \BibitemOpen
  \bibfield  {author} {\bibinfo {author} {\bibfnamefont {M.}~\bibnamefont
  {Shoda}}, \bibinfo {author} {\bibfnamefont {B.~D.}\ \bibnamefont
  {Chandran}},\ and\ \bibinfo {author} {\bibfnamefont {S.~R.}\ \bibnamefont
  {Cranmer}},\ }\href@noop {} {\bibfield  {journal} {\bibinfo  {journal} {arXiv
  preprint arXiv:2101.09529}\ } (\bibinfo {year} {2021})}\BibitemShut {NoStop}%
\bibitem [{\citenamefont {Kasper}\ \emph {et~al.}(2019)\citenamefont {Kasper},
  \citenamefont {Bale}, \citenamefont {Belcher}, \citenamefont {Berthomier},
  \citenamefont {Case}, \citenamefont {Chandran}, \citenamefont {Curtis},
  \citenamefont {Gallagher}, \citenamefont {Gary}, \citenamefont {Golub} \emph
  {et~al.}}]{kasper2019}%
  \BibitemOpen
  \bibfield  {author} {\bibinfo {author} {\bibfnamefont {J.}~\bibnamefont
  {Kasper}}, \bibinfo {author} {\bibfnamefont {S.}~\bibnamefont {Bale}},
  \bibinfo {author} {\bibfnamefont {J.~W.}\ \bibnamefont {Belcher}}, \bibinfo
  {author} {\bibfnamefont {M.}~\bibnamefont {Berthomier}}, \bibinfo {author}
  {\bibfnamefont {A.}~\bibnamefont {Case}}, \bibinfo {author} {\bibfnamefont
  {B.}~\bibnamefont {Chandran}}, \bibinfo {author} {\bibfnamefont
  {D.}~\bibnamefont {Curtis}}, \bibinfo {author} {\bibfnamefont
  {D.}~\bibnamefont {Gallagher}}, \bibinfo {author} {\bibfnamefont
  {S.}~\bibnamefont {Gary}}, \bibinfo {author} {\bibfnamefont {L.}~\bibnamefont
  {Golub}}, \emph {et~al.},\ }\href@noop {} {\bibfield  {journal} {\bibinfo
  {journal} {Nature}\ }\textbf {\bibinfo {volume} {576}},\ \bibinfo {pages}
  {228} (\bibinfo {year} {2019})}\BibitemShut {NoStop}%
\bibitem [{\citenamefont {Bale}\ \emph {et~al.}(2019)\citenamefont {Bale},
  \citenamefont {Badman}, \citenamefont {Bonnell}, \citenamefont {Bowen},
  \citenamefont {Burgess}, \citenamefont {Case}, \citenamefont {Cattell},
  \citenamefont {Chandran}, \citenamefont {Chaston}, \citenamefont {Chen} \emph
  {et~al.}}]{bale2019}%
  \BibitemOpen
  \bibfield  {author} {\bibinfo {author} {\bibfnamefont {S.}~\bibnamefont
  {Bale}}, \bibinfo {author} {\bibfnamefont {S.}~\bibnamefont {Badman}},
  \bibinfo {author} {\bibfnamefont {J.}~\bibnamefont {Bonnell}}, \bibinfo
  {author} {\bibfnamefont {T.}~\bibnamefont {Bowen}}, \bibinfo {author}
  {\bibfnamefont {D.}~\bibnamefont {Burgess}}, \bibinfo {author} {\bibfnamefont
  {A.}~\bibnamefont {Case}}, \bibinfo {author} {\bibfnamefont {C.}~\bibnamefont
  {Cattell}}, \bibinfo {author} {\bibfnamefont {B.}~\bibnamefont {Chandran}},
  \bibinfo {author} {\bibfnamefont {C.}~\bibnamefont {Chaston}}, \bibinfo
  {author} {\bibfnamefont {C.}~\bibnamefont {Chen}}, \emph {et~al.},\
  }\href@noop {} {\bibfield  {journal} {\bibinfo  {journal} {Nature}\ }\textbf
  {\bibinfo {volume} {576}},\ \bibinfo {pages} {237} (\bibinfo {year}
  {2019})}\BibitemShut {NoStop}%
\bibitem [{\citenamefont {Horbury}\ \emph {et~al.}(2020)\citenamefont
  {Horbury}, \citenamefont {Woolley}, \citenamefont {Laker}, \citenamefont
  {Matteini}, \citenamefont {Eastwood}, \citenamefont {Bale}, \citenamefont
  {Velli}, \citenamefont {Chandran}, \citenamefont {Phan}, \citenamefont
  {Raouafi} \emph {et~al.}}]{horbury2020}%
  \BibitemOpen
  \bibfield  {author} {\bibinfo {author} {\bibfnamefont {T.~S.}\ \bibnamefont
  {Horbury}}, \bibinfo {author} {\bibfnamefont {T.}~\bibnamefont {Woolley}},
  \bibinfo {author} {\bibfnamefont {R.}~\bibnamefont {Laker}}, \bibinfo
  {author} {\bibfnamefont {L.}~\bibnamefont {Matteini}}, \bibinfo {author}
  {\bibfnamefont {J.}~\bibnamefont {Eastwood}}, \bibinfo {author}
  {\bibfnamefont {S.~D.}\ \bibnamefont {Bale}}, \bibinfo {author}
  {\bibfnamefont {M.}~\bibnamefont {Velli}}, \bibinfo {author} {\bibfnamefont
  {B.~D.}\ \bibnamefont {Chandran}}, \bibinfo {author} {\bibfnamefont
  {T.}~\bibnamefont {Phan}}, \bibinfo {author} {\bibfnamefont {N.~E.}\
  \bibnamefont {Raouafi}}, \emph {et~al.},\ }\href@noop {} {\bibfield
  {journal} {\bibinfo  {journal} {Astrophys. J. Suppl. Ser.}\ }\textbf
  {\bibinfo {volume} {246}},\ \bibinfo {pages} {45} (\bibinfo {year}
  {2020})}\BibitemShut {NoStop}%
\bibitem [{\citenamefont {Tenerani}\ \emph {et~al.}(2020)\citenamefont
  {Tenerani}, \citenamefont {Velli}, \citenamefont {Matteini}, \citenamefont
  {R{\'e}ville}, \citenamefont {Shi}, \citenamefont {Bale}, \citenamefont
  {Kasper}, \citenamefont {Bonnell}, \citenamefont {Case}, \citenamefont
  {de~Wit} \emph {et~al.}}]{tenerani2020}%
  \BibitemOpen
  \bibfield  {author} {\bibinfo {author} {\bibfnamefont {A.}~\bibnamefont
  {Tenerani}}, \bibinfo {author} {\bibfnamefont {M.}~\bibnamefont {Velli}},
  \bibinfo {author} {\bibfnamefont {L.}~\bibnamefont {Matteini}}, \bibinfo
  {author} {\bibfnamefont {V.}~\bibnamefont {R{\'e}ville}}, \bibinfo {author}
  {\bibfnamefont {C.}~\bibnamefont {Shi}}, \bibinfo {author} {\bibfnamefont
  {S.~D.}\ \bibnamefont {Bale}}, \bibinfo {author} {\bibfnamefont {J.~C.}\
  \bibnamefont {Kasper}}, \bibinfo {author} {\bibfnamefont {J.~W.}\
  \bibnamefont {Bonnell}}, \bibinfo {author} {\bibfnamefont {A.~W.}\
  \bibnamefont {Case}}, \bibinfo {author} {\bibfnamefont {T.~D.}\ \bibnamefont
  {de~Wit}}, \emph {et~al.},\ }\href@noop {} {\bibfield  {journal} {\bibinfo
  {journal} {Astrophys. J. Suppl. Ser.}\ }\textbf {\bibinfo {volume} {246}},\
  \bibinfo {pages} {32} (\bibinfo {year} {2020})}\BibitemShut {NoStop}%
\bibitem [{\citenamefont {Howes}(2010)}]{howes2010}%
  \BibitemOpen
  \bibfield  {author} {\bibinfo {author} {\bibfnamefont {G.~G.}\ \bibnamefont
  {Howes}},\ }\href@noop {} {\bibfield  {journal} {\bibinfo  {journal} {MNRAS
  Lett.}\ }\textbf {\bibinfo {volume} {409}},\ \bibinfo {pages} {L104}
  (\bibinfo {year} {2010})}\BibitemShut {NoStop}%
\bibitem [{\citenamefont {Kawazura}\ \emph {et~al.}(2020)\citenamefont
  {Kawazura}, \citenamefont {Schekochihin}, \citenamefont {Barnes},
  \citenamefont {TenBarge}, \citenamefont {Tong}, \citenamefont {Klein},\ and\
  \citenamefont {Dorland}}]{kawazura2020}%
  \BibitemOpen
  \bibfield  {author} {\bibinfo {author} {\bibfnamefont {Y.}~\bibnamefont
  {Kawazura}}, \bibinfo {author} {\bibfnamefont {A.}~\bibnamefont
  {Schekochihin}}, \bibinfo {author} {\bibfnamefont {M.}~\bibnamefont
  {Barnes}}, \bibinfo {author} {\bibfnamefont {J.}~\bibnamefont {TenBarge}},
  \bibinfo {author} {\bibfnamefont {Y.}~\bibnamefont {Tong}}, \bibinfo {author}
  {\bibfnamefont {K.}~\bibnamefont {Klein}},\ and\ \bibinfo {author}
  {\bibfnamefont {W.}~\bibnamefont {Dorland}},\ }\href@noop {} {\bibfield
  {journal} {\bibinfo  {journal} {Phys. Rev. X}\ }\textbf {\bibinfo {volume}
  {10}},\ \bibinfo {pages} {041050} (\bibinfo {year} {2020})}\BibitemShut
  {NoStop}%
\bibitem [{\citenamefont {Greco}(1972)}]{greco1972}%
  \BibitemOpen
  \bibfield  {author} {\bibinfo {author} {\bibfnamefont {A.}~\bibnamefont
  {Greco}},\ }\href {https://www.osti.gov/biblio/4462654} {\bibfield  {journal}
  {\bibinfo  {journal} {Atti Accad. Naz. Lincei, Rend., Cl. Sci. Fis., Mat.
  Natur.}\ }\textbf {\bibinfo {volume} {52}} (\bibinfo {year}
  {1972})}\BibitemShut {NoStop}%
\bibitem [{\citenamefont {Anile}(1989)}]{anile1989}%
  \BibitemOpen
  \bibfield  {author} {\bibinfo {author} {\bibfnamefont {A.~M.}\ \bibnamefont
  {Anile}},\ }\href@noop {} {\emph {\bibinfo {title} {Relativistic fluids and
  magneto-fluids: With applications in astrophysics and plasma physics}}}\
  (\bibinfo  {publisher} {Cambridge University Press},\ \bibinfo {year}
  {1989})\BibitemShut {NoStop}%
\bibitem [{\citenamefont {Heyvaerts}\ \emph {et~al.}(2012)\citenamefont
  {Heyvaerts}, \citenamefont {Lehner},\ and\ \citenamefont
  {Mottez}}]{heyvaerts2012}%
  \BibitemOpen
  \bibfield  {author} {\bibinfo {author} {\bibfnamefont {J.}~\bibnamefont
  {Heyvaerts}}, \bibinfo {author} {\bibfnamefont {T.}~\bibnamefont {Lehner}},\
  and\ \bibinfo {author} {\bibfnamefont {F.}~\bibnamefont {Mottez}},\
  }\href@noop {} {\bibfield  {journal} {\bibinfo  {journal} {A\&A}\ }\textbf
  {\bibinfo {volume} {542}},\ \bibinfo {pages} {A128} (\bibinfo {year}
  {2012})}\BibitemShut {NoStop}%
\bibitem [{\citenamefont {Gammie}\ \emph {et~al.}(2003)\citenamefont {Gammie},
  \citenamefont {McKinney},\ and\ \citenamefont {T{\'o}th}}]{gammie2003}%
  \BibitemOpen
  \bibfield  {author} {\bibinfo {author} {\bibfnamefont {C.~F.}\ \bibnamefont
  {Gammie}}, \bibinfo {author} {\bibfnamefont {J.~C.}\ \bibnamefont
  {McKinney}},\ and\ \bibinfo {author} {\bibfnamefont {G.}~\bibnamefont
  {T{\'o}th}},\ }\href@noop {} {\bibfield  {journal} {\bibinfo  {journal}
  {Astrophys. J.}\ }\textbf {\bibinfo {volume} {589}},\ \bibinfo {pages} {444}
  (\bibinfo {year} {2003})}\BibitemShut {NoStop}%
\bibitem [{\citenamefont {Chandran}\ \emph {et~al.}(2018)\citenamefont
  {Chandran}, \citenamefont {Foucart},\ and\ \citenamefont
  {Tchekhovskoy}}]{chandran2018}%
  \BibitemOpen
  \bibfield  {author} {\bibinfo {author} {\bibfnamefont {B.~D.}\ \bibnamefont
  {Chandran}}, \bibinfo {author} {\bibfnamefont {F.}~\bibnamefont {Foucart}},\
  and\ \bibinfo {author} {\bibfnamefont {A.}~\bibnamefont {Tchekhovskoy}},\
  }\href@noop {} {\bibfield  {journal} {\bibinfo  {journal} {J. Plasma Phys.}\
  }\textbf {\bibinfo {volume} {84}} (\bibinfo {year} {2018})}\BibitemShut
  {NoStop}%
\bibitem [{\citenamefont {{Elsasser}}(1950)}]{elsasser1950}%
  \BibitemOpen
  \bibfield  {author} {\bibinfo {author} {\bibfnamefont {W.~M.}\ \bibnamefont
  {{Elsasser}}},\ }\href {https://doi.org/10.1103/PhysRev.79.183} {\bibfield
  {journal} {\bibinfo  {journal} {Phys. Rev.}\ }\textbf {\bibinfo {volume}
  {79}},\ \bibinfo {pages} {183} (\bibinfo {year} {1950})}\BibitemShut
  {NoStop}%
\bibitem [{Note1()}]{Note1}%
  \BibitemOpen
  \bibinfo {note} {In \protect \citet {chandran2018}, this was called the
  \uline {average fluid rest frame} (AFRF).}\BibitemShut {Stop}%
\bibitem [{\citenamefont {Matteini}\ \emph {et~al.}(2018)\citenamefont
  {Matteini}, \citenamefont {Stansby}, \citenamefont {Horbury},\ and\
  \citenamefont {Chen}}]{matteini2018}%
  \BibitemOpen
  \bibfield  {author} {\bibinfo {author} {\bibfnamefont {L.}~\bibnamefont
  {Matteini}}, \bibinfo {author} {\bibfnamefont {D.}~\bibnamefont {Stansby}},
  \bibinfo {author} {\bibfnamefont {T.}~\bibnamefont {Horbury}},\ and\ \bibinfo
  {author} {\bibfnamefont {C.~H.}\ \bibnamefont {Chen}},\ }\href@noop {}
  {\bibfield  {journal} {\bibinfo  {journal} {Astrophys. J. Lett.}\ }\textbf
  {\bibinfo {volume} {869}},\ \bibinfo {pages} {L32} (\bibinfo {year}
  {2018})}\BibitemShut {NoStop}%
\bibitem [{\citenamefont {Matteini}\ \emph {et~al.}(2015)\citenamefont
  {Matteini}, \citenamefont {Horbury}, \citenamefont {Pantellini},
  \citenamefont {Velli},\ and\ \citenamefont {Schwartz}}]{matteini2015}%
  \BibitemOpen
  \bibfield  {author} {\bibinfo {author} {\bibfnamefont {L.}~\bibnamefont
  {Matteini}}, \bibinfo {author} {\bibfnamefont {T.}~\bibnamefont {Horbury}},
  \bibinfo {author} {\bibfnamefont {F.}~\bibnamefont {Pantellini}}, \bibinfo
  {author} {\bibfnamefont {M.}~\bibnamefont {Velli}},\ and\ \bibinfo {author}
  {\bibfnamefont {S.}~\bibnamefont {Schwartz}},\ }\href@noop {} {\bibfield
  {journal} {\bibinfo  {journal} {Astrophys. J.}\ }\textbf {\bibinfo {volume}
  {802}},\ \bibinfo {pages} {11} (\bibinfo {year} {2015})}\BibitemShut
  {NoStop}%
\bibitem [{\citenamefont {Cranmer}\ and\ \citenamefont
  {Van~Ballegooijen}(2005)}]{cranmer2005}%
  \BibitemOpen
  \bibfield  {author} {\bibinfo {author} {\bibfnamefont {S.}~\bibnamefont
  {Cranmer}}\ and\ \bibinfo {author} {\bibfnamefont {A.}~\bibnamefont
  {Van~Ballegooijen}},\ }\href@noop {} {\bibfield  {journal} {\bibinfo
  {journal} {Astrophys. J. Suppl. Ser.}\ }\textbf {\bibinfo {volume} {156}},\
  \bibinfo {pages} {265} (\bibinfo {year} {2005})}\BibitemShut {NoStop}%
\bibitem [{\citenamefont {Verdini}\ and\ \citenamefont
  {Velli}(2007)}]{verdini2007}%
  \BibitemOpen
  \bibfield  {author} {\bibinfo {author} {\bibfnamefont {A.}~\bibnamefont
  {Verdini}}\ and\ \bibinfo {author} {\bibfnamefont {M.}~\bibnamefont
  {Velli}},\ }\href@noop {} {\bibfield  {journal} {\bibinfo  {journal}
  {Astrophys. J.}\ }\textbf {\bibinfo {volume} {662}},\ \bibinfo {pages} {669}
  (\bibinfo {year} {2007})}\BibitemShut {NoStop}%
\bibitem [{\citenamefont {Van~Ballegooijen}\ and\ \citenamefont
  {Asgari-Targhi}(2016)}]{ballegooijen2016}%
  \BibitemOpen
  \bibfield  {author} {\bibinfo {author} {\bibfnamefont {A.}~\bibnamefont
  {Van~Ballegooijen}}\ and\ \bibinfo {author} {\bibfnamefont {M.}~\bibnamefont
  {Asgari-Targhi}},\ }\href@noop {} {\bibfield  {journal} {\bibinfo  {journal}
  {Astrophys. J.}\ }\textbf {\bibinfo {volume} {821}},\ \bibinfo {pages} {106}
  (\bibinfo {year} {2016})}\BibitemShut {NoStop}%
\bibitem [{\citenamefont {van Ballegooijen}\ and\ \citenamefont
  {Asgari-Targhi}(2017)}]{ballegooijen2017}%
  \BibitemOpen
  \bibfield  {author} {\bibinfo {author} {\bibfnamefont {A.~A.}\ \bibnamefont
  {van Ballegooijen}}\ and\ \bibinfo {author} {\bibfnamefont {M.}~\bibnamefont
  {Asgari-Targhi}},\ }\href@noop {} {\bibfield  {journal} {\bibinfo  {journal}
  {Astrophys. J.}\ }\textbf {\bibinfo {volume} {835}},\ \bibinfo {pages} {10}
  (\bibinfo {year} {2017})}\BibitemShut {NoStop}%
\bibitem [{\citenamefont {{Perez}}\ and\ \citenamefont
  {{Chandran}}(2013)}]{perezchandran2013}%
  \BibitemOpen
  \bibfield  {author} {\bibinfo {author} {\bibfnamefont {J.~C.}\ \bibnamefont
  {{Perez}}}\ and\ \bibinfo {author} {\bibfnamefont {B.~D.~G.}\ \bibnamefont
  {{Chandran}}},\ }\href {https://doi.org/10.1088/0004-637X/776/2/124}
  {\bibfield  {journal} {\bibinfo  {journal} {Astrophys. J.}\ }\textbf
  {\bibinfo {volume} {776}},\ \bibinfo {eid} {124} (\bibinfo {year}
  {2013})}\BibitemShut {NoStop}%
\bibitem [{\citenamefont {Chandran}\ and\ \citenamefont
  {Perez}(2019)}]{chandran2019}%
  \BibitemOpen
  \bibfield  {author} {\bibinfo {author} {\bibfnamefont {B.~D.}\ \bibnamefont
  {Chandran}}\ and\ \bibinfo {author} {\bibfnamefont {J.~C.}\ \bibnamefont
  {Perez}},\ }\href@noop {} {\bibfield  {journal} {\bibinfo  {journal} {J.
  Plasma Phys.}\ }\textbf {\bibinfo {volume} {85}} (\bibinfo {year}
  {2019})}\BibitemShut {NoStop}%
\bibitem [{\citenamefont {Parker}(1965)}]{parker1965}%
  \BibitemOpen
  \bibfield  {author} {\bibinfo {author} {\bibfnamefont {E.}~\bibnamefont
  {Parker}},\ }\href@noop {} {\bibfield  {journal} {\bibinfo  {journal} {Space
  Sci. Rev.}\ }\textbf {\bibinfo {volume} {4}},\ \bibinfo {pages} {666}
  (\bibinfo {year} {1965})}\BibitemShut {NoStop}%
\bibitem [{\citenamefont {Hollweg}(1974)}]{hollweg1974}%
  \BibitemOpen
  \bibfield  {author} {\bibinfo {author} {\bibfnamefont {J.~V.}\ \bibnamefont
  {Hollweg}},\ }\href@noop {} {\bibfield  {journal} {\bibinfo  {journal} {J.
  Geophys. Res.}\ }\textbf {\bibinfo {volume} {79}},\ \bibinfo {pages} {1539}
  (\bibinfo {year} {1974})}\BibitemShut {NoStop}%
\bibitem [{\citenamefont {Akiyama}\ \emph
  {et~al.}(2021{\natexlab{a}})\citenamefont {Akiyama}, \citenamefont {Algaba},
  \citenamefont {Alberdi}, \citenamefont {Alef}, \citenamefont {Anantua},
  \citenamefont {Asada}, \citenamefont {Azulay}, \citenamefont {Baczko},
  \citenamefont {Ball}, \citenamefont {Balokovi{\'c}} \emph
  {et~al.}}]{akiyama2021}%
  \BibitemOpen
  \bibfield  {author} {\bibinfo {author} {\bibfnamefont {K.}~\bibnamefont
  {Akiyama}}, \bibinfo {author} {\bibfnamefont {J.~C.}\ \bibnamefont {Algaba}},
  \bibinfo {author} {\bibfnamefont {A.}~\bibnamefont {Alberdi}}, \bibinfo
  {author} {\bibfnamefont {W.}~\bibnamefont {Alef}}, \bibinfo {author}
  {\bibfnamefont {R.}~\bibnamefont {Anantua}}, \bibinfo {author} {\bibfnamefont
  {K.}~\bibnamefont {Asada}}, \bibinfo {author} {\bibfnamefont
  {R.}~\bibnamefont {Azulay}}, \bibinfo {author} {\bibfnamefont {A.-K.}\
  \bibnamefont {Baczko}}, \bibinfo {author} {\bibfnamefont {D.}~\bibnamefont
  {Ball}}, \bibinfo {author} {\bibfnamefont {M.}~\bibnamefont {Balokovi{\'c}}},
  \emph {et~al.},\ }\href@noop {} {\bibfield  {journal} {\bibinfo  {journal}
  {Astrophys. J. Lett.}\ }\textbf {\bibinfo {volume} {910}},\ \bibinfo {pages}
  {L12} (\bibinfo {year} {2021}{\natexlab{a}})}\BibitemShut {NoStop}%
\bibitem [{\citenamefont {Akiyama}\ \emph
  {et~al.}(2021{\natexlab{b}})\citenamefont {Akiyama}, \citenamefont {Algaba},
  \citenamefont {Alberdi}, \citenamefont {Alef}, \citenamefont {Anantua},
  \citenamefont {Asada}, \citenamefont {Azulay}, \citenamefont {Baczko},
  \citenamefont {Ball}, \citenamefont {Balokovi{\'c}} \emph
  {et~al.}}]{akiyama2021b}%
  \BibitemOpen
  \bibfield  {author} {\bibinfo {author} {\bibfnamefont {K.}~\bibnamefont
  {Akiyama}}, \bibinfo {author} {\bibfnamefont {J.~C.}\ \bibnamefont {Algaba}},
  \bibinfo {author} {\bibfnamefont {A.}~\bibnamefont {Alberdi}}, \bibinfo
  {author} {\bibfnamefont {W.}~\bibnamefont {Alef}}, \bibinfo {author}
  {\bibfnamefont {R.}~\bibnamefont {Anantua}}, \bibinfo {author} {\bibfnamefont
  {K.}~\bibnamefont {Asada}}, \bibinfo {author} {\bibfnamefont
  {R.}~\bibnamefont {Azulay}}, \bibinfo {author} {\bibfnamefont {A.-K.}\
  \bibnamefont {Baczko}}, \bibinfo {author} {\bibfnamefont {D.}~\bibnamefont
  {Ball}}, \bibinfo {author} {\bibfnamefont {M.}~\bibnamefont {Balokovi{\'c}}},
  \emph {et~al.},\ }\href@noop {} {\bibfield  {journal} {\bibinfo  {journal}
  {The Astrophysical Journal Letters}\ }\textbf {\bibinfo {volume} {910}},\
  \bibinfo {pages} {L13} (\bibinfo {year} {2021}{\natexlab{b}})}\BibitemShut
  {NoStop}%
\end{thebibliography}%
\end{document}